# Molecule dependent oxygen isotopic ratios in the coma of comet 67P/Churyumov-Gerasimenko

K. Altwegg[1*], H. Balsiger[1], M. Combi[2], J. De Keyser[3], M. N. Drozdovskaya[4], S. A. Fuselier[5,6], T. I. Gombosi[2], N. Hänni[1], M. Rubin[1], M. Schuhmann[1], I. Schroeder[1], S. Wampfler[4]

Affiliations:

[1]Physikalisches Institut, University of Bern, Sidlerstr. 5, CH-3012 Bern, Switzerland.

[2]Department of Climate and Space Sciences and Engineering, University of Michigan, 2455 Hayward, Ann Arbor, MI 48109, USA.

[3]Royal Belgian Institute for Space Aeronomy, BIRA-IASB, Ringlaan 3, B-1180 Brussels, Belgium.

[4]Center for Space and Habitability, University of Bern, Gesellschaftsstr. 6, CH-3012 Bern, Switzerland.

[5]Space Science Directorate, Southwest Research Institute, 6220 Culebra Rd., San Antonio, TX 78228, USA.

[6]Department of Physics and Astronomy, The University of Texas at San Antonio, San Antonio, TX, 78249

# Abstract

The ratios of the three stable oxygen isotopes $^{16}O$, $^{17}O$ and $^{18}O$ on Earth and, as far as we know in the solar system, show variations on the order of a few percent at most, with a few outliers in meteorites. However, in the interstellar medium there are some highly fractionated oxygen isotopic ratios in some specific molecules. The goal of this work is to investigate the oxygen isotopic ratios in different volatile molecules found in the coma of comet 67P/Churyumov-Gerasimenko and compare them with findings from interstellar clouds in order to assess commonalities and differences. To accomplish this goal, we analyzed data from the ROSINA instrument on Rosetta during its mission around the comet. $^{16}O/^{18}O$ ratios could be determined for $O_2$, methanol, formaldehyde, carbonyl sulfide and sulfur monoxide/dioxide. For $O_2$ the $^{16}O/^{17}O$ ratio is also available. Some ratios are strongly enriched in the heavy isotopes, especially for sulfur bearing molecules and formaldehyde, whereas for methanol the ratios are compatible with the ones in the solar system. $O_2$ falls in-between, but its oxygen isotopic ratios clearly differ from water, which likely rules out an origin of $O_2$ from water, be it by radiolysis, dismutation during sublimation or the Eley-Rideal process from water ions hitting the nucleus as postulated in the literature.

## 1. Introduction

Oxygen has three stable isotopes with mean terrestrial abundance ratios: $^{16}O/^{17}O$ = 2700, $^{16}O/^{18}O$ = 490 (Clayton, 2003). From meteoritic research, it is known that oxygen fractionates relatively moderately by few percent in the solar system with a mass-dependent relationship between variations in $^{16}O/^{17}O$ and $^{16}O/^{18}O$ ratios (e.g. Ireland et al., 2020). However, in the interstellar medium, large oxygen isotopic fractionations have been found in specific molecules (e.g. Loison et al., 2019). Isotopic ratios in different molecules can help to decipher their formation pathways including the physical boundary conditions. Many chemical reactions have different reaction rates for different isotopes, especially at very cold temperatures. This leads to fractionation with time. In comets, it was recognized already a long time ago, that for example the D/H ratio in water differs for most comets significantly from the terrestrial value. For heavier atoms, measurements of isotopic ratios in comets are scarce and normally come with big uncertainties. In the interstellar medium, such isotopic fractionation is even more pronounced, not only for hydrogen, but also for heavier atoms. The problem there is the difficulty to measure the main isotopologues as the star forming regions are very often optically thick for the more abundant species.

Thanks to the Rosetta mission and the ROSINA instrument (Balsiger et al., 2007), several isotopologues could be measured in the cometary coma of 67P/Churyumov-Gerasimenko with a relatively high precision. Apart from the deuterated species of $H_2O$ and $H_2S$ (Altwegg et al., 2015; Altwegg et al., 2016) this is especially true for the isotopologues of $CO_2$ (Hässig et al., 2017), of some sulfur bearing species (Calmonte et al., 2017), of silicon (Rubin et al., 2017), of some of the halogens (Dhooghe et al. 2017) and of oxygen in water (Schroeder et al., 2019). In addition most stable isotopes of the noble gases Xe (Marty et al., 2017), Kr and Ar (Rubin et al., 2018) have been detected in the cometary coma. Except for the isotopologues of $CO_2$ and probably the Ar isotopes, all isotopic ratios measured so far deviate to some degree from the terrestrial or solar values. This deviation clearly points to a presolar origin of most molecules detected in 67P and, in the case of Si and Xe, where chemistry cannot be the reason for fractionation, also to a non-homogenized solar nebula.

In this paper, we investigate some more oxygenated molecules in order to find out how the composition of 67P reflects the chemistry during presolar stages. Oxygen isotopic ratios can be derived for molecules which are not very rare in the coma of the comet and which have no strong mass interferences with neighboring molecules / fragments. Bulk abundances for such molecules can be found in Rubin et al. (2019). Relative abundances vary with heliocentric distance and latitude. Apart from the abundant water and carbon dioxide, $O_2$ with an abundance of (3.1±1.1) % relative to water, methanol (0.21±0.06 %), formaldehyde (0.3±0.1 %), OCS ($0.041_{-0.02}^{+0.08}$ %), SO ($0.071_{-0.037}^{+0.142}$ %) and $SO_2$ ($0.127_{-0.064}^{+0.254}$ %) are accessible for such isotopic studies.

One of the questions to answer is the formation pathway of $O_2$, which has a surprisingly high abundance (Bieler et al., 2015) and for which there exist several hypotheses on its formation. Loison et al. (2019) published a chemical model for the oxygen isotopic fractionation in the gas phase in cold clouds. This model yields high enrichments for the heavy oxygen isotopes in sulfur bearing molecules and $O_2$, but not in methanol and a depletion in formaldehyde. The question remains if this is also reflected in cometary ice.

## 2. Data

The ROSINA-DFMS mass spectrometer (Balsiger et al., 2007) flown on the Rosetta mission has a mass resolution m/Δm of ~3000 at the 1% level at *m/z* 28, which corresponds to roughly 9000 at FWHM. The dynamic range of one single spectrum for one integer *m/z* number with an integration time of 20 s is ~$10^4$. Depending on the species, this mass resolution and dynamic range limit the detection of the rare isotopologues. This is especially true for $^{17}O$ in carbon and sulfur bearing molecules due to interferences with the $^{13}C$ and $^{33}S$ isotopologues. However, the $^{16}O/^{18}O$ ratios in $CO_2$, $H_2O$, $O_2$, $H_2CO$, $CH_3OH$, OCS, SO/$SO_2$ are deduced by choosing special periods, where interference from neighboring species are minimal and abundances are maximal. In addition it is possible to deduce also $^{16}O/^{17}O$ in the abundant $H_2O$ and $O_2$. In the following we discuss the procedures and time range used for the different species. It has to be noted that due to electron impact ionization in the instrument, SO is a fragment of $SO_2$ and at the same time a parent. The same is true for CO and $CO_2$. Oxygen isotopic ratios for the monoxides are therefore not independent from the dioxides. All errors listed here or cited from earlier work represent 1-σ uncertainties.

DFMS peaks are fitted well by two Gaussians where the width of the second Gaussian is approximately 3 times the width and a tenth of the amplitude of the first (De Keyser et al., 2019). For a single spectrum, these widths are the same for all peaks, but they vary with mass and, due to temperatures effects, also slightly with time. If there is a large peak, these widths are fitted and the fit is applied to smaller peaks or overlapping peaks on the same mass. With this technique, it is possible to fit peaks even if they are mostly hidden inside larger peaks (Fig. 1). For spectra, where the amplitude is low, we omit the second Gaussian, as it is mainly hidden in the background.

The abundance measured by the mass spectrometer depends on the cometocentric distance, on the heliocentric distance and, due to the heterogeneous coma, also on the position of the spacecraft relative to the comet (northern/southern hemisphere). Early in the mission, between end September and end December 2014, the spacecraft was generally inside of 20 km from the comet. After this time, distances increased, partly due to special trajectories adapted to the needs of other instruments, partly due to the increased activity of the comet. In March 2016, the spacecraft again reached distances inside of 20 km, but only for a short period before going onto an extended tail excursion. In May 2016, the spacecraft was inside of 10 km of the nucleus. At that time ROSINA successfully targeted noble gases and by mid-August, when Rosetta was again close to the nucleus, the activity of the comet had decreased considerably, making the abundance of many of the molecules too low to detect rare isotopologues. This leaves the periods October-December 2014 and the short period in March 2016 as prime periods for this work.

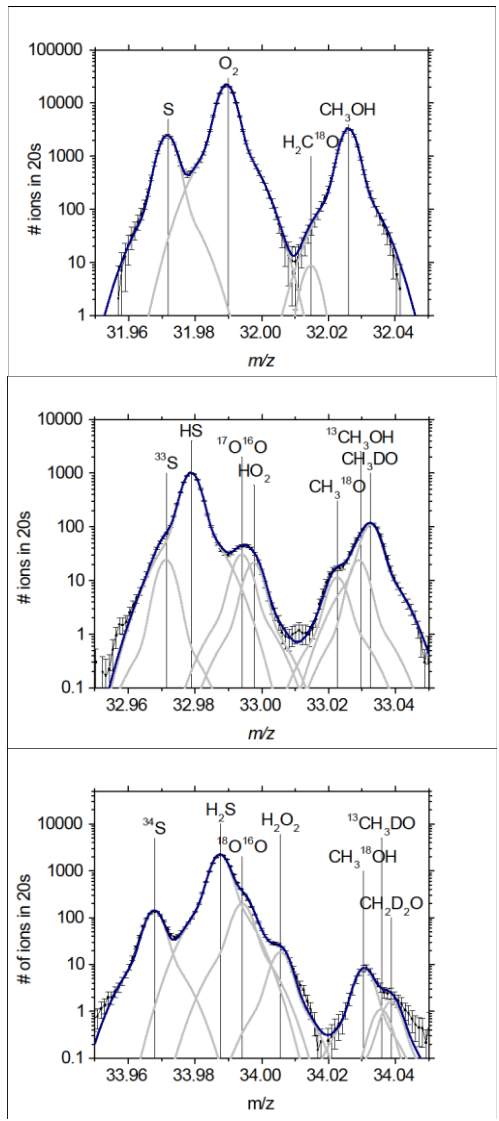

Fig. 1: Sample spectra used for the O$_2$ and CH$_3$OH isotopologues. Average of 7 spectra taken on Oct. 09, Oct. 17 and Oct. 19, 2014. Grey lines correspond to the individual fits with the blue line being the sum of all fits.

**CO / CO$_2$:** For CO there is a strong overlap of C$^{18}$O and NO and, as a result, a reliable value for $^{16}$O/$^{18}$O cannot be determined (Rubin et al., 2017). The same authors derived a value of 86.0 ± 8.5 for $^{12}$C/$^{13}$C. CO is a parent molecule but also a fragment of CO$_2$ from electron impact ionization and the isotopologues of CO and CO$_2$ are therefore not independent. Hässig et al. (2017) studied the CO$_2$ isotopologues in detail for two periods early in the mission: October 2014 and December 2014. There are several interferences, especially the interference between $^{17}$O and $^{13}$C isotopologues - true for all C-bearing molecules - which makes the detection of C$^{16}$O$^{17}$O impossible. The C$^{16}$O$^{18}$O peak might be contaminated by some NO$_2$, but this is probably not significant. This is discussed in detail by Hässig et al., 2017. Their main argument is a very good correlation between the main isotopologue of CO$_2$ on m/z 44 with the peak on m/z 46, independent of latitude or time period in the mission, which would not be expected if NO$_2$ would significantly interfere. The peak on *m/z* 46 is well resolved and a precise $^{16}$O/$^{18}$O ratio of 494 ± 8 was deduced (Hässig et al., 2017). They also derived a value of 84 ± 4 for the $^{12}$C/$^{13}$C ratio for CO$_2$, in agreement with the value derived for CO (Rubin et al., 2017).

**H$_2$O:** For water, the $^{18}$O isotopologue is well resolved on *m/z* 20, whereas H$_2^{17}$O on *m/z* 19 is part of a broad peak together with HDO. Schroeder et al., 2019 evaluated data for the entire mission duration for H$_2^{18}$O using also the fragments $^{16}$OH and $^{18}$OH, which are both well separated from neighboring molecules. H$_2^{17}$O needs manual fitting, $^{17}$OH is hidden by H$_2$O. The analysis of $^{17}$O was restricted to two periods, the first one in May 2015 during the inbound equinox, the second one in March 2016 during the second equinox. The first period was chosen as it represents probably best the bulk composition (see Rubin et al., 2019), the second one was a time period where the spacecraft was close to the nucleus and therefore the signal high enough. The evaluation of $^{16}$O/$^{18}$O shows, after correction for the detector gain due to ageing, that the ratio is constant over the mission. This is consistent with D/H in water, for which a very similar ratio was measured early as well as late in the mission (Altwegg et al. 2017). Although we cannot prove it for other species, it is very likely that the oxygen isotope ratios do not change over the mission in general, as mass dependent effects (e.g. sublimation rates) are probably very minor.

**O₂:** The abundance of O$_2$ follows largely the abundance of water (Bieler et al., 2015). The main isotopologue on *m/z* 32 is well separated from the neighboring S and CH$_3$OH. The $^{16}$O$^{17}$O molecule has an interference with HS and with HO$_2$, but can be reliably fitted. $^{16}$O$^{18}$O on *m/z* 34 is on the slope of the abundant H$_2$S peak. However, H$_2$S and O$_2$ do not follow each other (e.g. Läuter et al., 2020). Especially early in the mission, O$_2$ is by far the highest peak on *m/z* 32. To get a reliable value for $^{16}$O$^{18}$O, we therefore used periods where the abundance of O$_2$ is high enough to detect the $^{16}$O$^{17}$O and where, at the same time, the abundance of H$_2$S is relatively low. These conditions were fulfilled early in the mission. TableS1 (sup. mat.) gives the dates and times of the analyzed spectra and Fig. S1 (sup. mat.) shows the individually derived isotopic ratios. Fig. 1 shows sample spectra for *m/z* 32, 33 and 34 for this time period.

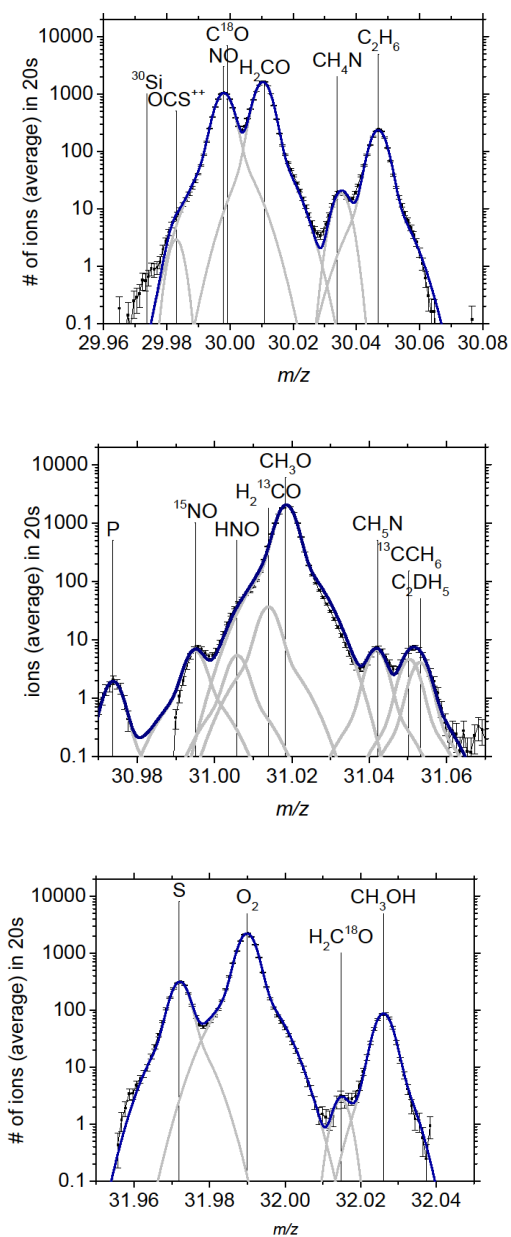

Fig. 2: Sample spectra used for the H$_2$CO isotopologues. Average of 12 spectra taken on Oct. 03 and 04, 2014 for *m/z* 30 and 32. For *m/z* 31, the spectrum shown is the average of 10 spectra taken on Dec. 10 (see text for an explanation of chosen periods). Grey lines correspond to the individual fits with the blue line being the sum of all fits.

**CH$_3$OH:** The main peak on *m/z* 32 is well separated from neighboring peaks. CH$_3$$^{17}$OH is hidden by $^{13}$CH$_3$OH. However, CH$_3$$^{18}$OH is well separated from neighboring peaks on *m/z* 34 (see Fig. 1), except for an interference with deuterated methanol isotopologues $^{13}$CH$_3$DO and CH$_2$D$_2$O, which are clearly less abundant. Deuterated methanol will be the topic of a forthcoming paper. Count rates were never high enough to deduce $^{16}$O/$^{18}$O from single spectra. Therefore, in this case spectra were co-added before the peaks were fitted, using the same spectra as for O$_2$. The resulting $^{16}$O/$^{18}$O ratio was deduced from the mean value over three days, for which each 7 spectra were co-added (Oct. 09, Oct. 17 and Oct. 19, 2014). We also derived the $^{12}$C/$^{13}$C ratio for methanol (Fig. 1). $^{13}$CH$_3$OH has some overlap with deuterated methanol, but is reliably fitted for times when methanol is high and statistics therefore not a problem, which are the same periods as used for O$_2$.

**H$_2$CO**: Analysis of formaldehyde is more difficult. The main peak on *m/z* 30 is clearly separated from C$^{18}$O and NO, which strongly overlap, and CH$_4$N and C$_2$H$_6$. However, formaldehyde is never very abundant. In addition, H$_2$CO is also a fragment of methanol. Fragmentation patterns measured for methanol for DFMS show that the ratio of the fragment on mass 30 to the parent methanol ion is ~0.1 (Schuhmann et al., 2019). H$_2$C$^{18}$O is in the slope of methanol, which has a similar bulk abundance as formaldehyde and the isotopologue also interferes with the much more abundant $O_2$ (see Fig. 2), mainly because of the limited dynamic range of single spectra. For the $^{16}O/^{18}O$ in formaldehyde, we need periods when the $O_2$ abundance was relatively low, the signal for $H_2CO$ was as high as possible, and the ratio methanol/formaldehyde as low as possible. The only time when this was fulfilled was early October 2014. At that time methanol was about 10 times smaller than the peak on m/z 30 associated with $H_2CO$. That means the contribution of the methanol fragments to the peaks of formaldehyde is on the order of 1% and is therefore negligible. We analyzed data from Oct. 03 and 04, 2014 (Fig. 2). Here the $H_2C^{18}O$ peak is visible.

We also tried to derive the $^{12}C/^{13}C$ ratio for formaldehyde (Fig. 2). Because $H_2^{13}CO$ is hidden by the peak of $CH_3O$, a fragment of methanol, we looked for a period when we had a very strong formaldehyde signal, a high ratio of formaldehyde to methanol, independent of $O_2$. A period was found in Dec. 2014, where $H_2CO$ was high, but so was $O_2$. Because of the high $O_2$, $H_2C^{18}O$ cannot be reliably fitted during this period because of the limited dynamic range (Fig. 1). While fitting *m/z* 30 and 31 for $H_2^{13}CO/H_2^{12}CO$ with data from the December period, we also derived as a by-product the $^{12}C_2H_6$ /$^{13}C^{12}CH_6$ ratio and from there the $^{12}C/^{13}C$ ratio in ethane. In the same spectra there is also the $^{15}NO$ isotopologue. Nitrogen isotopes will be the topic of another paper. For ethane we deduced a $^{12}C/^{13}C$ ratio of 105 ± 10, slightly higher than the value of $^{12}C/^{13}C$ = 85.5 ± 9.0 deduced by Rubin et al., 2017 for the fragment $C_2H_5$. The difference is mainly due to a better understanding of the ageing of the detector (De Keyser et al., 2019).

**OCS:** For all sulfur bearing species, there is a strong interference between the sulfur isotopologues and the oxygen isotopologues. However, the mass differences are such that, at least for $^{18}O$ and $^{34}S$, two separate peaks are fitted for the DFMS mass spectra. The abundance of OCS was highest in March 2016. The main peak is well separated from oxygenated hydrocarbons. For the $^{18}O$ isotopologue, there is a weak interference from $CH_2OS$ (see Fig. 3). 10 spectra from March 16/17, 2016 were co-added before fitting the peaks and deriving the $^{16}O/^{18}O$ and the $^{32}S/^{34}S$ ratios

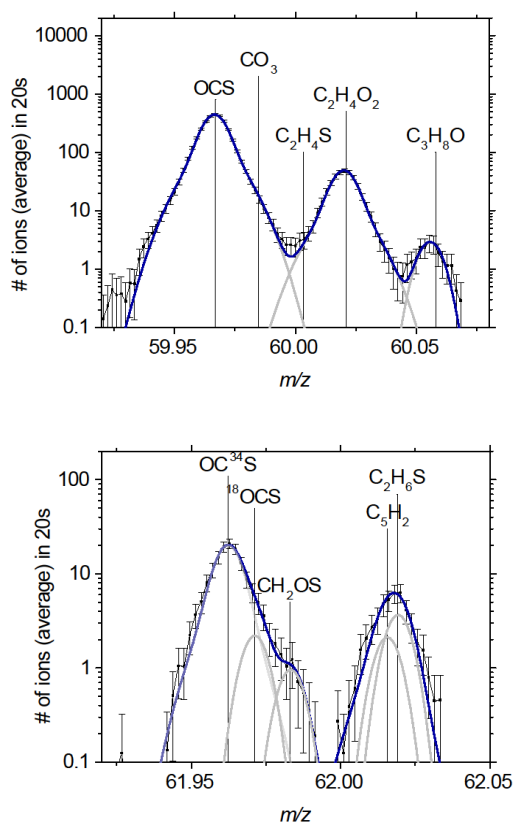

Fig. 3: Sample spectra for OCS, average of 10 individual spectra taken on March 16, 2016. Grey lines correspond to the individual fits with the blue line being the sum of all fits.

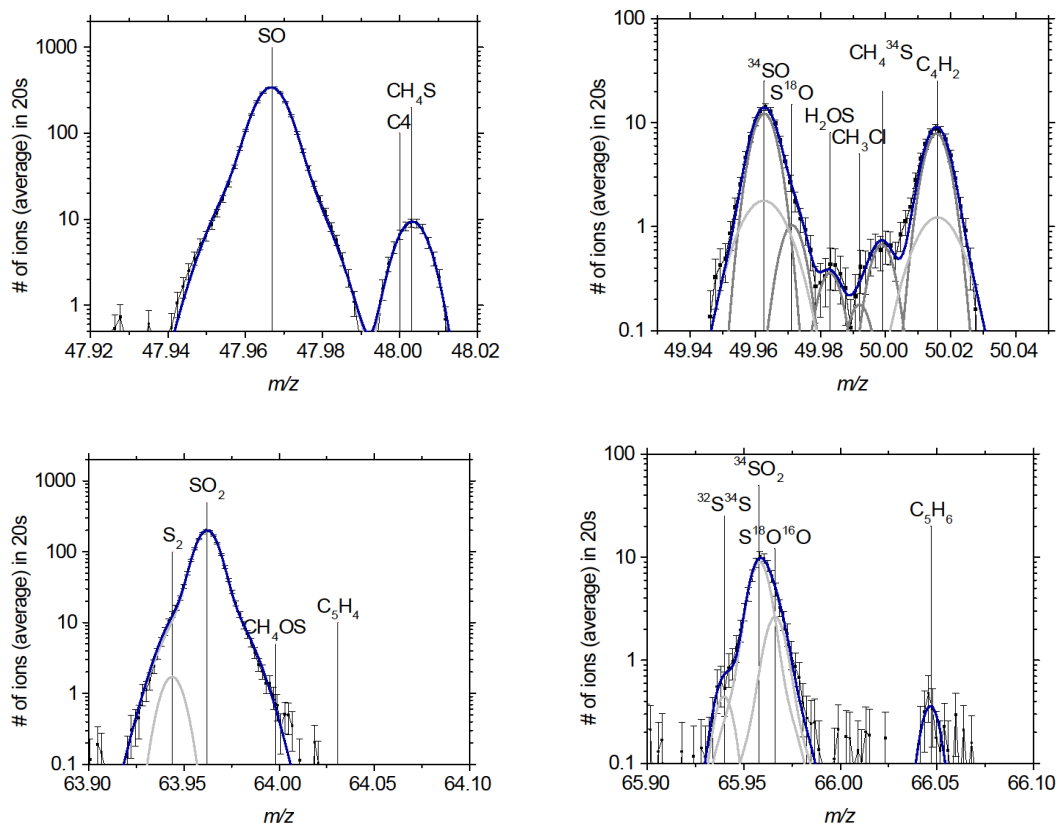

Fig. 4: Sample spectra for SO and SO$_2$. Shown are averaged spectra from 12 individual spectra on Dec. 10, 2014. Grey lines correspond to the individual fits with the blue line being the sum of all fits.

**SO/SO$_2$:** SO is a fragment of SO$_2$ from electron impact ionization in the instrument and at the same time also a parent molecule (Calmonte et al., 2016). The isotopologues of the two species are therefore not independent. While SO is well separated from any interfering peak, SO$_2$ interferes with S$_2$, on *m/z* 64 as well as on *m/z* 66 (Fig. 4). However, early in the mission SO$_2$ was clearly dominant (Calmonte et al. 2016). Again, the sulfur and oxygen isotopologues are close together in mass. For SO$_2$, the signal for the oxygen isotopologue on *m/z* 66 compared to the sulfur isotopologue is twice as high as for SO due to the two oxygen atoms, but the instrument sensitivity for the *m/z* range 60-66 is only about 64% compared to *m/z* 48. The highest abundances were measured on Dec. 09/10 2014. 12 spectra on each day were co-added before fitting the peaks. The two values were then averaged to get the $^{16}O/^{18}O$ ratio for SO and SO$_2$ as well as the $^{32}S/^{34}S$ ratio.

**Uncertainties:** While abundances of the main isotopologues are generally well determined with little statistical error, this is not the case for the minor isotopologues. The only species where the uncertainty is low (1.6 %) is $CO_2$ (Hässig et al., 2017). For $H_2O$, the error is governed by the uncertainty in the detector gain (see Schroeder et al., 2019). This gain is automatically set by DFMS according to the signal strength and is therefore for the main isotopologue on *m/z* 18 generally significantly different from the gain for the minor ones. This leads to a 1-σ uncertainty of ~8 % for the $^{16}O/^{18}O$ ratio. For the other species, the gain is mostly on the highest level for all isotopologues, which means its uncertainty cancels out. The main uncertainty therefore is due to statistics and fitting errors. This is especially true for $H_2CO$, where the number of usable spectra is very limited. For the other species the uncertainty of the $^{16}O/^{18}O$ ratio is 10 - 30 %.

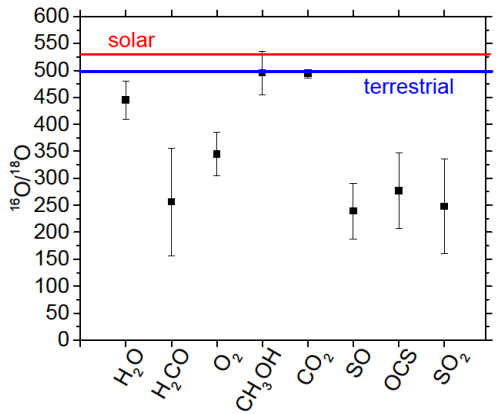

Fig. 5: $^{16}O/^{18}O$ ratios for 8 different molecules ordered by mass. $CO_2$ from Hässig et al. (2017); $H_2O$ from Schroeder et al. (2019); all other molecules from this work. The red line denotes the solar value, the blue line the terrestrial.

## 3. Results

The ROSINA data to date allowed the derivation of the $^{16}O/^{18}O$ ratio for eight different volatile molecules in the coma of 67P. For two of them, the ratio $^{16}O/^{17}O$ is available. For the sulfur bearing molecules the sulfur isotope ratio $^{32}S/^{34}S$ was derived from the exact same dataset, as there is an interference between sulfur and oxygen isotopes. This allows a direct comparison of these ratios, which show strong deviations from terrestrial in oxygen but not sulfur. We were able to also derive some carbon isotopic ratios at the same time. All results with their uncertainties are given in table 1 and the $^{16}O/^{18}O$ ratios also in Fig. 5.

From the table and Fig. 5 it is clear that, albeit the error bars, variations between oxygen isotopic fractionation are quite large. On one end, there are ratios in $CO_2$ and methanol, which are compatible with terrestrial values, at the other end there are ratios in all sulfur bearing molecules, formaldehyde and $O_2$, which are enriched in $^{18}O$ by a factor 2. Water is somewhere in-between. Sulfur isotopes do not vary very much and are all, within their uncertainties, compatible with the standard value from the Vienna-Canyon Diablo Troilite (VCDT) (Robinson et al., 1995). Surprisingly, carbon isotopic fractionation displays some large differences between species.

Table 1: Oxygen, carbon and sulfur isotopic ratios with 1-σ errors.

| Species | $^{16}O/^{18}O$ (terrestrial 498.7[*]/solar 530[**]) | $^{16}O/^{17}O$ (terrestrial 2632[*] /solar 2798[**]) | $^{32}S/^{34}S$ (Standard solar system 22.64[***]) | $^{12}C/^{13}C$ (terrestrial 92.4[****] / solar 88.88[*****]) | Ref. * Clayton, 2003 ** Mc Keegan et al., 2011 ***Robinson, 1995 ****Boutton, 1991 |
|---|---|---|---|---|---|

| | | | | *****Meibom et al., 2008 |
|---|---|---|---|---|
| $H_2O$ | 445 ± 35 | 2182 ± 170 | | Schroeder et al., 2019 |
| $O_2$ | 345 ± 40 | 1544 ± 308 | | this work |
| $CO_2$ | 494 ± 8 | | 84 ± 4 | Hässig et al., 2017 |
| CO | | | 86 ± 8 | Rubin et al., 2017 |
| $CH_3OH$ | 495 ± 40 | | 91 ± 10 | this work |
| $H_2CO$ | 256 ± 100 | | 40 ± 14 | this work |
| SO | 239 ± 52 | 23.5 ± 2.5 | | this work |
| $SO_2$ | 248 ± 88 | 21.3 ± 2.1 | | this work |
| OCS | 277 ± 70 | 21.7 ± 4.0 | | this work |
| $C_2H_6$ | | | 105± 10 | this work |

## 4. Discussion

**$H_2O$:** Water is enriched relative to terrestrial values in the heavy isotopologues by (17 ± 6) % and (11 ± 7) % for $^{17}O$ and $^{18}O$ (Schroeder et al., 2019). It has been shown by the same authors that this value is compatible with self-shielding models which predict a 5-20% enrichment (Lyons & Young, 2005; Young, 2007; Lee et al., 2008). Values measured remotely in different comets show a diverse picture. But due to the large uncertainties, they are mostly compatible with the value derived for 67P as well as with the terrestrial value (see Schroeder et al., 2019 and references therein). It is difficult to conclude that self-shielding is really the cause for this fractionation. Other molecules are much more fractionated than water and probably need another explanation, which then may also be applicable to water.

**$O_2$**: Oxygen is enriched in the heavy isotopes by (41 ± 18) % and (30 ± 7) % for $^{17}O$ and $^{18}O$ relative to terrestrial. The relative enrichment of $^{18}O$ to $^{17}O$ is similar to that of water with $^{18}O/^{17}O$ = 4.5 ± 1.0 for $O_2$ and 4.9 ± 0.5 for water, respectively. While error bars are rather large, it is nevertheless clear that the isotopic fractionation in $O_2$ for both heavy oxygen isotopes is not compatible with water. The incompatibility with water immediately rules out the proposed mechanisms based on water for the origin of molecular oxygen in 67P, like radiolysis from water (Mousis et al., 2016) and dismutation of $H_2O_2$ (Dulieu et al., 2017). Apart from other arguments (e.g. Heritier et al., 2018), it also rules out the Eley-Rideal scenario proposed by Yao et al. (2017), especially as the oxygen isotopic ratio found in the dust of 67P is close to solar (Paquette et al., 2018). Interaction between energetic water ions and the surface of 67P is very unlikely to yield highly fractionated $O_2$. This leaves a primordial origin of $O_2$ as described in Taquet et al. (2016) for grain-surface chemistry or gas phase chemistry as described in Rawlings et al. (2019).

**CO, $CO_2$, $H_2CO$, $CH_3OH$**: $CO_2$ in 67P has a terrestrial value (Hässig et al., 2016) for oxygen isotopes, but it is not compatible with the value derived from solar wind for the Sun. Methanol is compatible with the terrestrial fractionation, but, within error bars, also with the solar value for $^{16}O/^{18}O$. Chemically, methanol is probably a product from CO by successive hydrogenation on cold grains (e.g. Watanabe & Kouchi, 2002). In this case, it would make sense that $CO_2$ and methanol share the same oxygen isotopic fractionation, but only if $CO_2$ is the product of oxygenation of CO from an isotopically solar/terrestrial like reservoir. Possible scenarios on the interconnection between CO and $CO_2$ in comets are discussed in A'Hearn et al. (2012). The strong mass interference of $C^{18}O$ and NO and the fact that CO is not only a parent, but also a fragment of $CO_2$ from electron impact ionization make it impossible to derive a meaningful value for the $^{16}O/^{18}O$ in the parent CO from ROSINA. In order to assess better the situation between $CO_2$ and $CH_3OH$, we also derived the $^{12}C/^{13}C$ ratio (see Fig.1) for methanol. This ratio is 84 ± 4 for $CO_2$ as derived by Hässig et al. (2017) and 86 ± 8 for CO (Rubin et al., 2017). For methanol we determine a ratio of 91 ± 10, which overlaps within error bars with the ratios for CO and $CO_2$. This strengthens the formation pathway of methanol and $CO_2$, both starting with CO.

One would expect that $H_2CO$ and $CH_3OH$ also share the same isotopic fractionation for C as well as for O if most of the $H_2CO$ is formed by the same mechanism as methanol, namely by successive hydrogenation of CO. However, our result with $^{16}O/^{18}O$ = 256 ± 100 clearly has a substantial enrichment of $^{18}O$ for formaldehyde, yet this enrichment is not seen for methanol. For formaldehyde deriving a value for $H_2^{13}CO$ is challenging, but the evaluation of spectra measured over two days gave very consistent results with $H_2^{12}CO/H_2^{13}CO$ = 40 ± 14, clearly lower than for $CO_2$ or methanol. This strengthens the notion, that $H_2CO$ and $CH_3OH$ do not share their formation pathway. The ratio of 6 ± 2 for $H_2^{13}CO/H_2C^{18}O$ is compatible with solar and with most measured interstellar values (e.g. Kutner et al., 1982; Henkel et al., 1979). This ratio therefore cannot be used as argument that oxygen and carbon isotopic ratios are solar.

Measured $^{12}C/^{13}C$ ratios in the interstellar medium yield a somewhat heterogeneous picture. In a review article Wilson & Rood (1994) claim that the $^{12}C/^{13}C$ ratio shows a clear trend with distance to the galactic center. The values are generally higher for $H_2CO$ than for CO. A recent paper by Persson et al. (2018) has solid numbers for $^{12}C/^{13}C$ and $^{18}O/^{16}O$ in formaldehyde towards IRAS 16293–2422 B, a low mass star forming region. Unfortunately there are no values for methanol. It is, however, interesting to note that in this case $^{12}C/^{13}C$=56, that is significantly lower than solar while $^{16}O/^{18}O$ is around 800, significantly higher than solar for formaldehyde.

Wirström et al., (2011; 2012) have measured strong differences in the carbon isotopic fractionation for CO and $CH_3OH$ compared to $H_2CO$ in massive young stellar objects (YSOs). They derived $^{12}C/^{13}C$ ratios for CO and $CH_3OH$ in different YSOs, demonstrating that in most cases the ratios were equal for the two molecules within error bars, although they differ from source to source. The picture is very different for $H_2CO$, where the ratio deviates significantly from the one of CO in the same source. This clearly points to a different formation mechanism for the two molecules. It is argued that, in principle, $^{12}C/^{13}C$ is expected to be higher for molecules formed in the gas phase as CO freezes out onto cold grains with a preference for the heavy isotopes, which then leaves CO in the gas phase enriched in $^{12}C$ and also $^{16}O$. The results by Wirström et al. (2012) for e.g. the source DR21 (OH) with a $^{12}C/^{13}C$ ratio of 58 ± 8 for CO and 55 ± 8 for methanol as opposed to ~18 for formaldehyde shows the contrary. Even if these massive YSOs are probably not good proxies for our solar system, our results with different oxygen and carbon isotopic fractionation for methanol and formaldehyde are very much in line with these observations. The heterogeneous picture for formaldehyde and methanol may reflect the fact that part of $H_2CO$ is formed by hydrogenation along with $CH_3OH$, but another part may also be a product of gas phase chemistry (van der Tak et al., 2000). However, there are unfortunately little data for methanol and formaldehyde for the same sources measured with the same technique / telescope.

**Sulfur bearing oxygenated molecules**: All oxygenated sulfur bearing molecules share a similar oxygen isotopic fractionation with a depletion of $^{16}O/^{18}O$ of almost a factor of 2 relative to terrestrial. Their ratios are very similar to that for $H_2CO$. On the other hand, the sulfur isotopic ratios are, within error bars, compatible with the VCDT standard for sulfur (Robinson, 1995). For the sulfur species $H_2S$, $CS_2$ and OCS in 67P Calmonte et al. (2017) found a slight depletion in the heavy sulfur isotopes. A depletion in the heavy sulfur isotopes cannot be confirmed or ruled out for the oxygenated species considered here, as the error bars are too large. The average $^{34}S/^{18}O$ is 11.4 ± 3.3 for the sulfur bearing oxygenated species. There are values for the OCS minor isotopologues in source B of IRAS 16293–2422 (Drozdovskaya et al., 2018). For this source $^{34}$S/ $^{18}$O is 20 and $^{13}$C/$^{18}$O = 10. The uncertainties of these ratios are ~30%. As the main isotopologue is optically thick in source B of IRAS 16293–2422, there are no meaningful ratios for sulfur or oxygen isotopes in this survey. However, a survey towards Orion KL of carbon-sulfur species by Tercero et al. (2010) gives $^{16}$O/$^{18}$O ratios for OCS of 200 ± 112 and 250 ± 135 for the extended ridge and the plateau, respectively. These values are very much in accordance with our measurements, even though Orion KL may again not be a good proxy for our solar system.

SO and $SO_2$ measured by ROSINA are not independent as SO is a fragment of $SO_2$ in the mass spectrometer. Not surprisingly, their oxygen isotopic ratios are very similar with an enrichment of the $^{18}$O by a factor 2 relative to solar. For SO, there are data for four different dense clouds/cores by Loison et al. (2019) for the $^{16}$O/$^{18}$O ratios. They range from 70 to 170, i.e., even more enriched in heavy isotopes than our values.

Loison et al. (2019) also modelled the oxygen fractionation as a function of time in cold, dense clouds. They found a large $^{18}$O enrichment in the gas phase for OCS, SO, $SO_2$, and $O_2$ after about $10^5$ y, but a depletion for $H_2CO$, mostly due to isotopic exchange reactions. Water (OH) is slightly enriched in their model, methanol stays almost constant with time. Our results show a very similar trend except for formaldehyde. Even if this model is for the gas phase of cold clouds at 10K and therefore not directly applicable for the ice in comets, the mechanisms at play may be very similar. The problem with formaldehyde, as outlined above, is not yet understood.

# 5. Conclusion

During the Rosetta mission we found conditions favorable to study oxygen isotopologues for different molecules in the coma of 67P. While the isotopologues of water and $CO_2$ have been published earlier by Schroeder et al. (2019) and by Haessig et al. (2017), respectively, in this paper we deduced $^{16}$O/$^{17}$O and $^{16}$O/$^{18}$O for $O_2$, and $^{16}$O/$^{18}$O for methanol, formaldehyde, carbonyl sulfide and sulfur monoxide/dioxide. Except for $CO_2$ and methanol, all heavy isotopologues are enriched relative to terrestrial / solar values, but to different degrees. Water is enriched by less than 20% in both heavy isotopologues relative to solar, while all the other species are enriched by roughly a factor 2. For some of the species we also looked at the carbon isotopologues and found an enrichment in $^{13}$C for formaldehyde, while the values for CO (Rubin et al., 2017), $CO_2$ (Hässig et al., 2017) and methanol are compatible with terrestrial.

Our values seem to fit well the (relatively scarce) values measured in the ISM, although most of these measurements were done in massive cores, which may not be representative for the origin of the cometary material. The model by Loison et al. (2019) for the enrichment of $^{18}$O in the gas phase in a cold cloud may also not be representative, but it is likely that the reactions used in this model, especially the isotopic transfer reactions, were also at play when the material formed which now makes up comet 67P. The riddle about the formation of methanol and formaldehyde has been pointed out before. While it is very likely that methanol formed from CO on dust grains by hydrogenation as they share their oxygen and carbon isotopic ratios, formaldehyde has probably followed another pathway of formation, seen from the different carbon and oxygen isotopologues in the ISM (Wirström et al., 2012) as well as in comet 67P.

The isotopic ratios of $O_2$ found in 67P, which do not match the water isotopic ratios, rule out several of the proposed formation scenarios of $O_2$ and, at least for the moment, point to a primordial origin of this quite abundant species.

# Acknowledgements

ROSINA would not have produced such outstanding results without the work of the many engineers, technicians, and scientists involved in the mission, in the Rosetta spacecraft team, and in the ROSINA instrument team over the last 20 years, whose contributions are gratefully acknowledged. Rosetta is an ESA mission with contributions from its member states and NASA. We acknowledge herewith the work of the whole ESA Rosetta team. We also would like to thank the anonymous reviewer for the very helpful comments. Work at the University of Bern was funded by the State of Bern, the Swiss National Science Foundation (SNSF, 200020_182418), the Swiss State Secretariat for Education, Research and Innovation (SERI) under contract number 16.0008- 2, and by the European Space Agency's PRODEX Program. SFW acknowledges the financial support of the SNSF Eccellenza Professorial Fellowship PCEFP2_181150. JDK acknowledges support by the Belgian Science Policy Office via PRODEX/ROSINA PEA 90020. SAF acknowledges JPL contract 1496541. Work at UoM was supported by contracts JPL 1266313 and JPL 1266314 from the US Rosetta Project. MND is supported by the Swiss National Science Foundation (SNSF) Ambizione grant 180079, the Center for Space and Habitability (CSH) Fellowship, and the IAU Gruber Foundation Fellowship.

**Data availability**: *The data underlying this article are available from the ESA Rosetta data archive at* https://www.cosmos.esa.int/web/psa/rosetta *and its mirror site at NASA Small Bodies Node* https://pds-smallbodies.astro.umd.edu/data_sb/missions/rosetta/index.shtml

# Supplementary material: Anomalous oxygen isotopic ratios

Table S1: date and times of analyzed O$_2$ isotopologues. Times given correspond to *m/z* 32.

| | | | |
|---|---|---|---|
| 1  | Packet time: 10/09/2014 15:32:18.114 | 17 | Packet time: 11/28/2014 08:23:46.159 |
| 2  | Packet time: 10/09/2014 16:15:35.552 | 18 | Packet time: 12/03/2014 15:00:42.392 |
| 3  | Packet time: 10/09/2014 16:57:59.009 | 19 | Packet time: 12/08/2014 06:36:49.501 |
| 4  | Packet time: 10/09/2014 17:41:22.570 | 20 | Packet time: 12/09/2014 00:40:21.130 |
| 5  | Packet time: 11/10/2014 01:13:40.351 | 21 | Packet time: 12/09/2014 01:31:41.165 |
| 6  | Packet time: 11/10/2014 02:05:57.843 | 22 | Packet time: 12/09/2014 02:14:08.636 |
| 7  | Packet time: 11/28/2014 00:39:06.099 | 23 | Packet time: 12/09/2014 07:58:34.582 |
| 8  | Packet time: 11/28/2014 01:36:36.460 | 24 | Packet time: 12/09/2014 14:08:02.220 |
| 9  | Packet time: 11/28/2014 02:24:46.223 | 25 | Packet time: 12/09/2014 14:59:42.234 |
| 10 | Packet time: 11/28/2014 03:12:47.727 | 26 | Packet time: 12/09/2014 15:42:46.032 |
| 11 | Packet time: 11/28/2014 03:59:43.454 | 27 | Packet time: 12/09/2014 16:25:49.898 |
| 12 | Packet time: 11/28/2014 04:48:41.297 | 28 | Packet time: 12/09/2014 17:08:47.819 |
| 13 | Packet time: 11/28/2014 05:30:59.130 | 29 | Packet time: 12/09/2014 17:51:49.580 |
| 14 | Packet time: 11/28/2014 06:14:46.864 | 30 | Packet time: 12/09/2014 19:18:28.813 |
| 15 | Packet time: 11/28/2014 06:57:36.670 | 31 | Packet time: 12/09/2014 20:01:28.588 |
| 16 | Packet time: 11/28/2014 07:40:40.397 | | |

Figure S1: Individual results for the oxygen isotopic ratios in O$_2$. Red: terrestrial value, blue: solar value. Upper panel $^{16}$O/$^{18}$O, lower panel $^{16}$O/$^{17}$O.

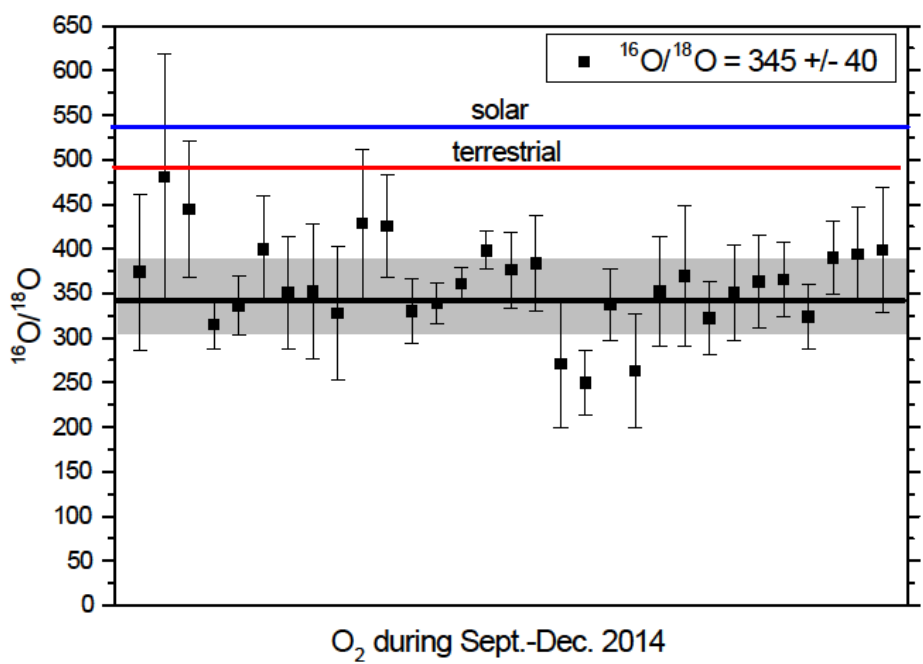

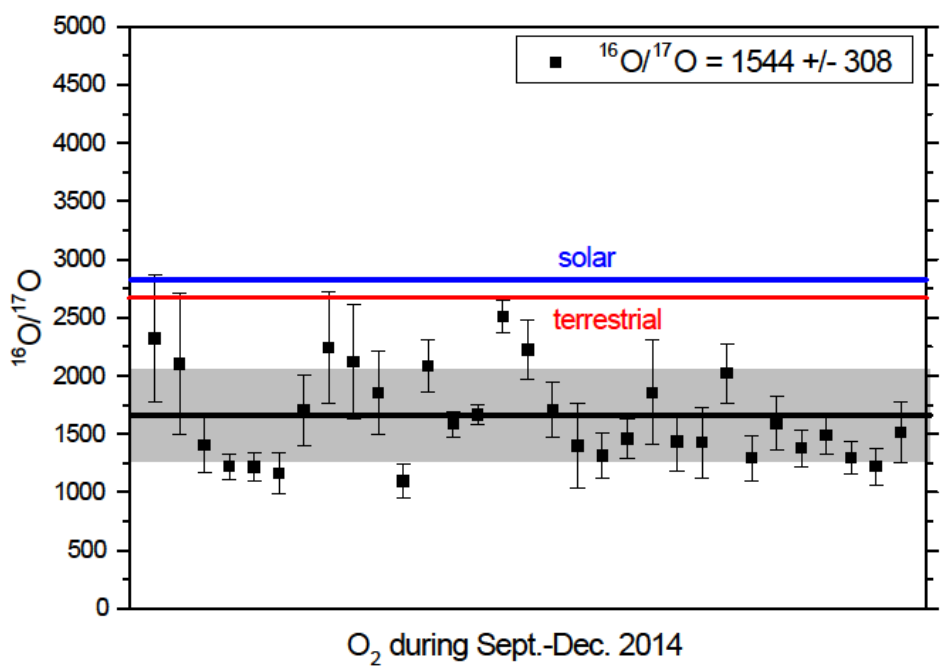

$O_2$ during Sept.-Dec. 2014